\documentstyle[12pt,epsfig]{article}

\newtheorem{optimizationproblem}{Optimization Problem}

\newcommand{\x}{\vec{x}} 
\newcommand{\y}{y}       
\newcommand{\Sample}{{S}}  
\newcommand{\n}{n}       
\newcommand{\Learner}{{\cal{L}}}    
\newcommand{\AL}{\vec{\alpha}}   
\newcommand{\w}{\vec{w}}   

\newcommand{\sgn}{\hbox{sgn}}     

\newcommand{\fig}[1]{Fig.~\protect\ref{#1}}

\newcommand{\svmlight}{SVM$^{\rm Light}$}

\title{Mapping Subsets of Scholarly Information}
\author{Paul Ginsparg\footnote{Presenter of invited talk at Arthur M. Sackler Colloquium on "Mapping Knowledge Domains", 9--11 May 2003,
Beckman Center of the National Academy of Sciences, Irvine, CA} ,
Paul Houle, Thorsten Joachims, and Jae-Hoon  Sul\\
(Cornell University, Ithaca, NY 14853, USA)}

\begin{document}
\maketitle
\begin{abstract} 
We illustrate the use of machine learning techniques to 
analyze, structure, maintain, and evolve a large online corpus of academic literature.
An emerging field of research can be identified as part of an existing corpus, permitting the implementation of a more coherent community structure for its practitioners.
\end{abstract}

\section{Background} \label{sec:bg}

The arXiv\footnote{See http://arXiv.org/ . For general background, see
 \cite{Ginsparg/01}.}
is an automated repository of over 250,000 full-text research
articles\footnote{as of mid-Oct 2003}
in physics and related disciplines, going back over a decade and growing at a
rate of 40,000 new submissions per year. It serves over 10 million requests per
month \cite{Ginsparg/03}, including tens of thousands of search queries per
day, and over 20 million full text downloads during calendar year '02.
It is a significant example of a Web-based service that
 has changed the practice of research in
a major scientific discipline.  It provides nearly comprehensive coverage
of large areas of physics, and serves as an on-line seminar system for
those areas. 
It also provides a significant resource for model building and algorithmic experiments in mapping
scholarly domains. Together with the SLAC SPIRES-HEP database\footnote{The  Stanford Linear Accelerator Center SPIRES-HEP database has comprehensively catalogued the High Energy Particle Physics (HEP) literature online since 1974, and indexes more than 500,000 high-energy physics related articles including their full citation tree (see \cite{O'Connell:2000uc}).}, it provides a public resource of full-text articles and associated citation tree of many millions of links, with a focused disciplinary coverage, and rich usage data.

In what follows, we will use arXiv data to illustrate how machine learning methods can be used to analyze, structure, maintain, and evolve a large online corpus of academic literature.
The specific application will be to train a support vector machine text
classifier to extract an emerging research area from a larger-scale resource.
The automated detection of such subunits can play an
important role in disentangling other sub-networks and associated
sub-communities from the global network.  Our notion of ``mapping'' here
is in the mathematical sense of associating attributes to objects, rather
than in the sense of visualization tools. While the former underlies the latter,
we expect the two will increasingly go hand-in-hand
(for a recent review, see \cite{Borner/etal/03}).

\section{Text Classification} \label{sec:tcat}

The goal of text classification is the automatic assignment of
documents to a fixed number of semantic categories. In the
 ``multi-label'' setting, each document can
be in zero or one or more categories. Efficient automated techniques are essential to avoid tedious and expensive manual category assigment for large document sets.
A ``knowledge engineering'' approach, involving hand-crafting accurate text
classification rules, is surprisingly difficult and time-consuming
\cite{Hayes/Weinstein/90}. We therefore take a machine learning approach to
generating text classification rules automatically from examples.

The machine learning approach can be phrased as a supervised
learning problem. The learning task is represented by the training sample $\Sample_n$
\begin{eqnarray}
(\x_1,\y_1), (\x_2,\y_2),\ldots, (\x_\n,\y_\n)
\end{eqnarray}
of size $\n$ documents, where $\x_i$ represents the
document content. In the multi-label setting, each
category label is treated as a separate binary classification problem. For
each such binary task, $\y_i \in \left\{-1,+1\right\}$ indicates
whether a document belongs to a particular class. The task of the
learning algorithm $\Learner$ is to find a decision rule $h_\Learner: \x
\longrightarrow \left\{-1,+1\right\}$ based on $\Sample_n$ that
classifies new documents $\x$ as accurately as possible.

\begin{figure}[tp]
\begin{center} \leavevmode
\psfig{figure=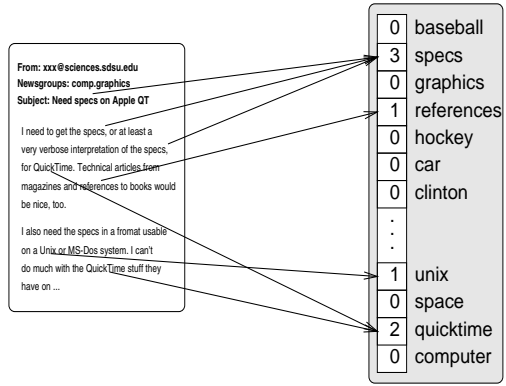,height=2.6in,width=2.0in,angle=270}
\caption{Representing text as a feature vector.}
\label{fig:attribval}  
\end{center}
\end{figure}

Documents need to be transformed into a representation suitable
for the learning algorithm and the classification task.
Information Retrieval research suggests
that words work well as representation units, and that for many tasks
their ordering can be ignored without losing too much information.
This type of representation is commonly called the ``bag-of-words''
model, an attribute-value representation of text.
Each text document is represented by a vector in the lexicon space, i.e.,
by a ``term frequency'' feature vector TF$(w_i,x)$, with component values equal
to the number of times each distinct word $w_i$ 
in the corpus occurs in the document $x$.
\fig{fig:attribval} shows an example feature vector for a
particular document. 

This basic representation is ordinarily refined in a few ways:
\begin{description}
\item[TF$\times$IDF Weighting:] Scaling the components
  of the feature vector with their {\em inverse document frequency}
  IDF$(w_i)$ \cite{Salton/Buckley/88} often leads to improved
  performance. In general, IDF$(w_i)$ is some decreasing function of the word
  frequency DF$(w_i)$, equal to the number of documents in the corpus which contain
  the word $w_i$. For example,
  \begin{eqnarray}
  {\rm IDF}(w_i) = \log\left(\frac{n}{{\rm DF}(w_i)}\right) \label{eq:idf}
  \end{eqnarray}
  where $n$ is the total number of documents. Intuitively, the inverse
  document frequency assumes that rarer terms have more significance for classification 
  purposes, and hence gives them greater weight. To compensate for the effect
  of different document lengths, each document feature vector $\vec{x}_i$
  is normalized to unit length: $||\vec{x}_i||=1$.

\item[Stemming:] Instead of treating each occurrence form of a word as a different
  feature, stemming is used to project the different forms of a
  word onto a single feature, the word stem, by removing inflection
  information \cite{Porter/80}. For example ``computes'',
  ``computing'', and ``computer'' are all mapped to the same stem
  ``comput''. The terms ``word'' and ``word stem'' will be used
  synonymously in the following.
  
\item[Stopword Removal:] For many classification tasks, common words
  like ``the'', ``and'', or ``he'' do not help discriminate between document
  classes. Stopword removal describes the process of eliminating such
  words from the document by matching against a predefined list of
  stop-words. We use a standard stoplist of roughly 300 words.
\end{description}

\section{Support Vector Machines} \label{sec:svm}

SVMs \cite{Vapnik/98a} were developed by V. Vapnik et al.\ based on the
structural risk minimization principle from statistical learning
theory. They have proven to be a highly effective method for learning
text classification rules, achieving state-of-the-art performance on a
broad range of tasks \cite{Joachims/98a,Dumais/etal/98}. Two main
advantages of using SVMs for text classification lie in their ability
to handle the high dimensional feature spaces arising from the
bag-of-words representation. From a statistical perspective, they are
robust to overfitting and are well suited for the statistical
properties of text. 
From a computational perspective, they can be trained efficiently despite the
large number of features. A detailed overview of the SVM approach to text
classification, with more details on the notation used below,
is given in \cite{Joachims/02a}.

\begin{figure}[t]
\begin{center} \leavevmode
\psfig{figure=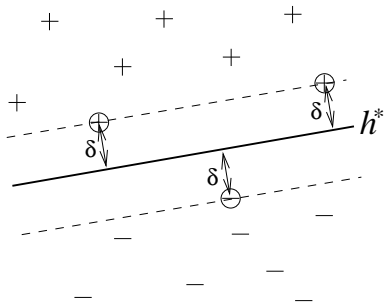,height=4cm,angle=0}
\caption{A binary classification problem ($+$ vs. $-$) in two
  dimensions.  The hyperplane $h^{*}$ separates positive and negative training examples with maximum margin $\delta$. The examples closest to the hyperplane are called {\em support vectors} (marked with circles).} \label{fig:svm_ex}
\end{center} 
\end{figure}

In their basic form, SVMs learn linear decision rules 
\begin{eqnarray}
  h(\x) = \sgn\{ \w \cdot \x + b\}\ ,\label{eq:svmldr}
\end{eqnarray} 
described by a weight vector $\w$ and a threshold $b$, from an input
sample of $\n$ training examples $\Sample_n=((\x_1,\y_1), \cdots,
(\x_\n,\y_\n))$, $\x_i \in {\bf I\kern-.18em R}^N$, $\y_i \in \{-1,+1\}$. For a
linearly separable $\Sample_n$, the SVM finds the hyperplane with
maximum Euclidean distance to the closest training examples. This
distance is called the margin $\delta$, as depicted in
\fig{fig:svm_ex}. Geometrically, the hyperplane is defined by its normal
vector, $\w$, and its distance from the origin, $-b$.
For non-separable training sets, the amount of
training error is measured using slack variables $\xi_i$.

Computing the position of the hyperplane is equivalent to solving the following
convex quadratic optimization problem \cite{Vapnik/98a}:
\begin{optimizationproblem}[SVM (primal)] \label{op:svmsoftprimal}
\vspace*{-0.06cm}
\begin{eqnarray}
\mbox{\rm minimize: } & & V(\w,b,\vec{\xi}) = \frac{1}{2}\w\cdot\w + C \: \sum_{i=1}^{\n}\xi_i \label{eq:softobjindsvmprimal} \\ 
\mbox{\rm subj.\ to: } & & \forall_{i=1}^{\n}: \y_i [\w \cdot \x_i + b] \ge 1-\xi_i \label{eq:softconstindsvmprimal} \\
                      & & \forall_{i=1}^{\n}: \xi_i>0 
\end{eqnarray} 
\end{optimizationproblem}
The margin of the resulting hyperplane is $\delta=1 / ||\w||$.

The constraints (\ref{eq:softconstindsvmprimal}) require that all
training examples are classified correctly up to some slack $\xi_i$.
If a training example lies on the ``wrong'' side of the hyperplane, we have
the corresponding $\xi_i\ge1$, and thus
$\sum_{i=1}^{\n}\xi_i$ is an upper bound on the number of training
errors. The factor $C$ in (\ref{eq:softobjindsvmprimal}) is a
parameter that allows trading off training error vs.\ model complexity.
The optimal value of this parameter depends on the particular
classification task and must be chosen via cross-validation or by
some other model selection strategy. For text classification, however,
the default value of $C={1}/{{\rm max}_i ||\x_i||^2}=1$ has proven to
be effective over a large range of tasks \cite{Joachims/02a}.

OP\ref{op:svmsoftprimal} has an equivalent dual formulation:

\begin{optimizationproblem}[SVM (dual)] \label{op:svmsoftdual}
\begin{eqnarray}
\mbox{\rm maximize: \hspace{-0.7cm}} & & W\!(\AL) = \!\sum_{i=1}^{\n}\alpha_i - \frac{1}{2}\!\sum_{i=1}^{\n}\sum_{j=1}^{\n}\y_i \y_j \alpha_i \alpha_j (\x_i\cdot\x_j) \label{eq:softobjindsvmdual}\\
\mbox{\rm subj.\ to: \hspace{-0.7cm}} & & \sum_{i=1}^{\n}\y_i \alpha_i = 0 \label{tj:eq:softconsteqindsvmdual}\\
& & \forall {i \in [1..\n]}: 0 \le \alpha_i \le C \label{eq:softconstindsvmdual}
\end{eqnarray}
\end{optimizationproblem}

From the solution of the dual, the classification rule solution can be
constructed as
\begin{eqnarray} 
\w  = \sum_{i=1}^{\n}\alpha_i \y_i \x_i
& \mbox{\hspace*{0.01cm} and \hspace*{0.01cm}} 
& b = \y_{usv} - \w \! \cdot \! \x_{usv} \ ,\label{eq:svmdecrule} 
\end{eqnarray}
where $(\x_{usv},\y_{usv})$ is some training example with $0 <
\alpha_{usv} < C$.  For the experiments in this paper, \svmlight\
\cite{Joachims/02a} is used for solving the dual optimization
problem\footnote{\svmlight\ is available at
http://svmlight.joachims.org/}. More 
detailed introductions to SVMs can be found in
\cite{Burges/98a,Cristianini/Shawe-Taylor/00}.

An alternative to the approach here would be to use Latent Semantic Analysis (LSA/LSI)
\cite{Landauer/Dumais/97} to generate the feature vectors.
LSA can potentially give better recall by capturing partial synonymy in the word
representation, i.e., by bundling related words into a single feature.
The reduction in dimension can also be computationally efficient, facilitating
capture of the full text content of papers, including their citation information.
On the other hand, the SVM already scales well to a large feature space, and 
performs well at determining relevant content through suitable choices of weights.
On a sufficiently large training set, the SVM might even
benefit from access to fine distinctions between words potentially obscured
by the LSA bundling.  It is thus an open question, well worth further investigation, as to whether
LSA applied to full text could improve performance of the SVM in this application, without loss
of computational efficiency.
Generative models for text classification provide yet another alternative approach,
in which  each document is viewed as a mixture of topics, as in the 
statistical inference algorithm for Latent Dirichlet Allocation
used in \cite{Griffiths/Steyvers/03}. This approach can in principle provide significant
additional information about document content, but does not scale well to a large document
corpus, so the real-time applications intended here would not yet
be computationally feasible in that framework.

\section{arXiv Benchmarks} 

Before using the machine learning framework to identify new
subject area content, we first assessed its performance on
the existing (author-provided) category classifications. Roughly 180,000 titles
and abstracts were fed to model building software which constructed a
lexicon of roughly 100,000 distinct words
and produced training files containing the TD$\times$IDF document vectors
for \svmlight.  (While the SVM machinery could easily be used to 
analyze the full document content, previous experiments \cite{Joachims/02a} 
suggest that well-written titles and abstracts provide a highly focused
characterization of content at least as effective for our document
classification purposes.) The set of support vectors and weight
parameters output by \svmlight\ was converted into a form specific
to the linear SVM, eq.~(\ref{eq:svmldr}): a weight vector $\w_c$
and a threshold $b_c$, where $c$ is an index over the categories.

\begin{figure}[tp]
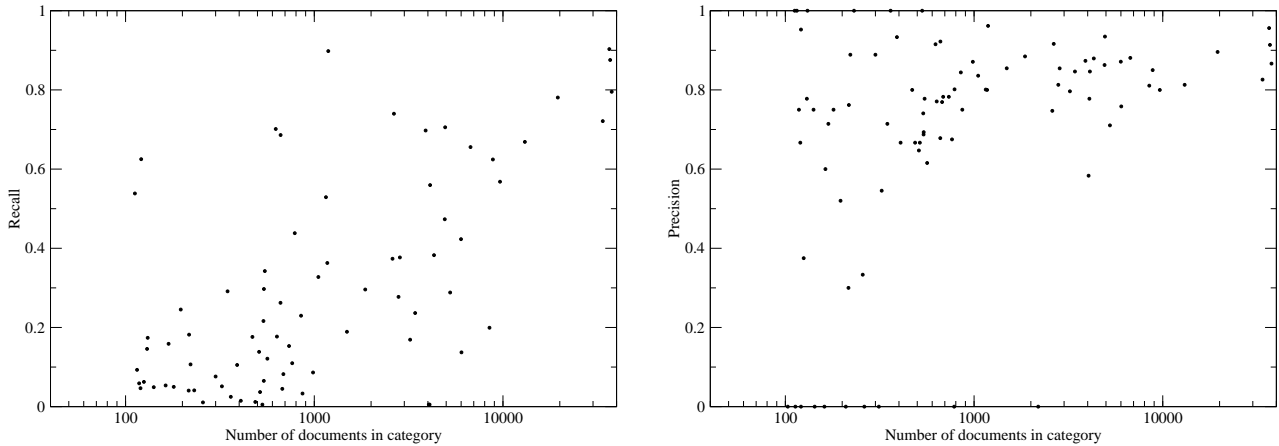

\begin{center} \leavevmode
\hglue-.2in\psfig{figure=images/recall.eps,width=3.2in}
\hglue.2in\psfig{figure=images/precision.eps,width=3.2in}
\caption{Recall and Precision as functions of category size for 78 arXiv major
categories and minor subject classes.  2/3 of the sample was used as a training
set and 1/3 as a test set. The four largest categories, each with over 30,000
documents are cond-mat, astro-ph, hep-th, and hep-ph.}
\label{fig:rfcs}
\end{center}
\end{figure}

As seen in \fig{fig:rfcs},  the success of the SVM in classifying
documents improves as the size of a category increases.  The SVM is remarkably
successful at identifying documents in large ($>10,000$ documents) categories
and less successful on smaller subject areas ($< 500$ documents).  A cutoff was
imposed to exclude subject areas with fewer than 100 documents.\footnote{Some
of the smaller subject areas are known to be less topically focused, so the
difficulty in recall, based solely on title/abstract terminology, was
expected.}

In experiments with small categories $(N<1000)$, the SVM consistently chose
thresholds that resulted in unacceptably low recall (i.e., missed documents
relevant to the category).  This is in part because the null hypothesis, that
no documents are in the category, describes the data well for a small category:
e.g., if only 1000 of 100,000 documents are in a category, the null hypothesis
has a 99\% accuracy.\footnote{``Accuracy'' is the percentage of documents correctly classified.
We also employ other common terminology from information retrieval:
``precision'' is the fraction of those documents retrieved that relevant to a query, and ``recall'' 
is the fraction of all relevant documents in the collection that are retrieved.}
To counteract the bias against small categories, 10\% of
the training data was used as a validation set to fit a sigmoid probability
distribution 1/(1+exp(Af+B)) \cite{Platt/99}. This converts the dot product
$\x_i\cdot\w_c$ between document vector and weight vector into a probability
$P(i \in c\;|\; x_i)$ that
document $i$ is a member of category $c$, given its feature vector $x_i$.
Use of the probability in place of
the uncalibrated signed distance from the hyperplane output by the SVM
permitted greater levels of recall for small categories.

Other experiments showed that the use of TF$\times$IDF weighting as in
eq.~(\ref{eq:idf}) improved accuracy consistently over pure TF weighting, so
TF$\times$IDF weighting was used in the experiments to follow. We also used
a document frequency threshhold to exclude rare words from the lexicon, but
found little difference in accuracy between using a document occurrence
threshold of two and five.\footnote{Words that appeared in fewer than two
documents constituted roughly 50\% of the lexicon, and those that appeared in
fewer than five documents roughly 70\%. Ignoring rare and consequently
uninformative words hence reduces the computational needs.}
Increasing the weight of title words with respect to abstract words, on the
other hand, consistently worsened accuracy, indicating that words in a
well-written abstract contain as much or more content classification import as
words
in the title. Changes in the word tokenization and stemming algorithms did not
have a significant impact on overall accuracy. 
The default value of $C=1$ in eq.~(\ref{eq:softconstindsvmdual}) was preferred.

\section{q-bio Extraction} 

There has been recent anecdotal evidence of an intellectual trend among physicists towards work in biology, ranging from biomolecules, molecular pathways and networks, gene expression, cellular and multicellular systems to population dynamics and evolution. This work has appeared in separate parts of the archive, particularly
under ``Soft Condensed Matter'', ``Statistical Mechanics'', ``Disordered Systems and Neural Networks'', ``Biophysics, and ``Adaptive and Self-Organizing Systems''
(abbreviated cond-mat.soft, cond-mat.stat-mech, cond-mat.dis-nn, physics.bio-ph, and
nlin.AO).
A more coherent forum for the exchange of these ideas was requested, under the nomenclature ``Quantitative Biology'' (abbreviated ``q-bio'').

To identify first whether there was indeed a real trend to nurture and amplify, 
and to create a training set, volunteers were enlisted to identify the q-bio content from the above subject areas in which it was most highly focused.  Of 5565 such articles received from Jan 2002 through Mar 2003, 466 (8.4\%) were found to have one of the above biological topics as its primary focus.  The total number of distinct words in these 
titles, abstracts, plus author names, was 23558, of which 7984 were above the ${\rm DF} = 2$ document frequency threshold. (Author names were included in the analysis since they have potential ``semantic'' content in this context, i.e., are potentially useful for document classification. The SVM algorithm will automatically determine whether or not to use the information by choosing suitable weights.)

A data-cleaning procedure was employed, in which \svmlight\ was first run with
$C=10$.  We recall from eqs.~(\ref{eq:softobjindsvmprimal}) and
(\ref{eq:softconstindsvmdual}) that larger $C$ penalizes training errors and
requires larger $\alpha$'s to fit the data. Inspecting the ``outlier''
documents \cite{Guyon/etal/96} with the largest $|\alpha_i|$ then permitted
manual cleaning of the training set. 10 were moved into q-bio, and 15 moved
out, for a net movement to 461 q-bio (8.3\%) of the 5565 total.  Some of the others
flagged involved word confusions, e.g., ``genetic algorithms''
typically involved programming rather than biological applications. Other
``q-bio'' words with frequent non-biological senses were ``epidemic'' (used for
rumor propagation), ``evolution'' (used also for dynamics of sandpiles),
``survival probabilities'', and extinction.  ``Scale-free networks'' were
sometimes used for biological applications, and sometimes not.
To help resolve some of these ambiguities, the vocabulary
was enlarged to include a list of most frequently used two-word phrases with
semantic content different from their constituent words .

\begin{figure}[tp]
\begin{center} \leavevmode
\psfig{figure=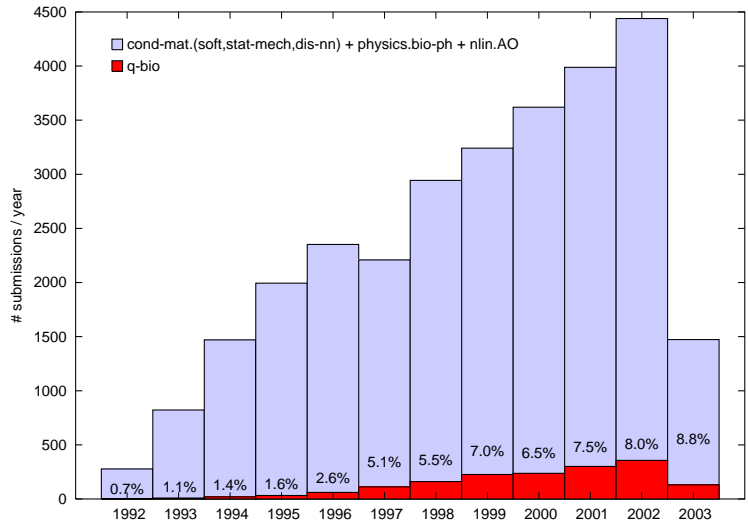,width=4.0in}
\caption{The number of submissions per year from 1992 through April 2003 in
particular subsets of arXiv subject areas of cond-mat, physics, and nlin most
likely to have
``quantitative biology'' content. The percentage of q-bio content in these
areas grew from roughly 1\% to nearly 10\% during this timeframe, suggesting
a change in intellectual activity among arXiv-using members of these
communities.}
\label{fig:qbio}
\end{center}
\end{figure}

\begin{figure}[tp]
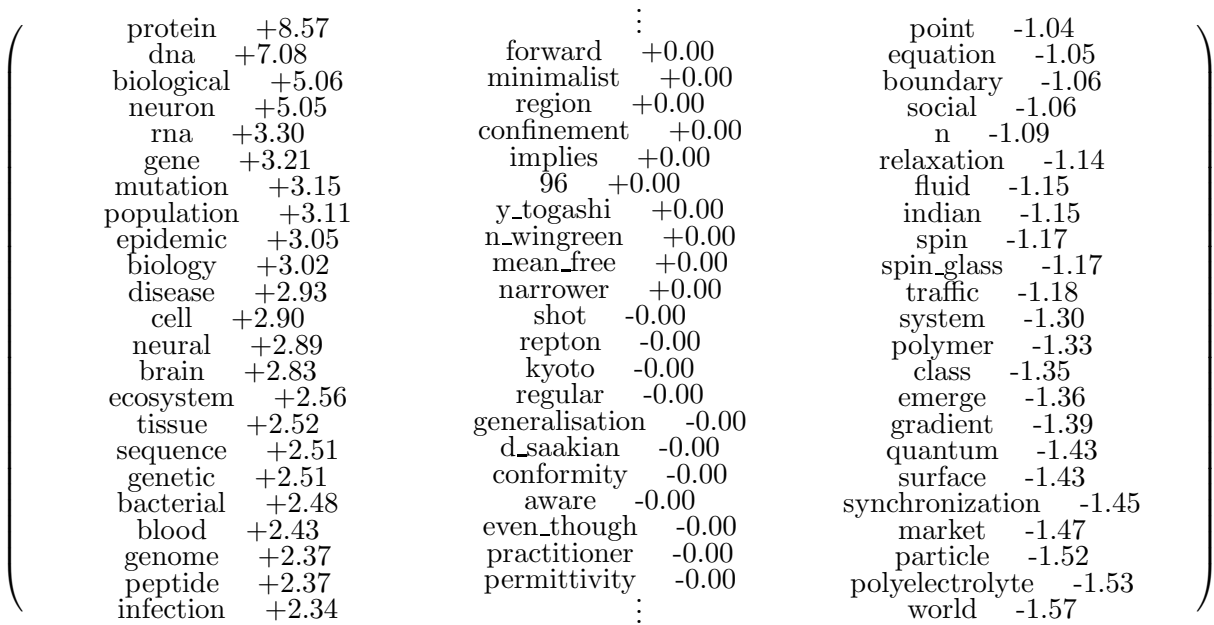

\begin{center} \leavevmode
\raise1.53in\hbox{$\left(\vbox{\vskip1.6in}\right.$}
\hbox{\small\baselineskip10pt\hsize2in\vbox{\obeylines
protein \quad+8.57
dna \quad+7.08
biological \quad+5.06
neuron \quad+5.05
rna \quad+3.30
gene \quad+3.21
mutation \quad+3.15
population \quad+3.11
epidemic \quad+3.05
biology \quad+3.02
disease \quad+2.93
cell \quad+2.90
neural \quad+2.89
brain \quad+2.83
ecosystem \quad+2.56
tissue \quad+2.52
sequence \quad+2.51
genetic \quad+2.51
bacterial \quad+2.48
blood \quad+2.43
genome \quad+2.37
peptide \quad+2.37
infection \quad+2.34}\vbox{\obeylines
 \qquad $\vdots$
forward \quad+0.00
minimalist \quad+0.00
region \quad+0.00
confinement \quad+0.00
implies \quad+0.00
96 \quad+0.00
y\_togashi \quad+0.00
n\_wingreen \quad+0.00
mean\_free \quad+0.00
narrower \quad+0.00
shot \quad-0.00
repton \quad-0.00
kyoto \quad-0.00
regular \quad-0.00
generalisation \quad-0.00
d\_saakian \quad-0.00
conformity \quad-0.00
aware \quad-0.00
even\_though \quad-0.00
practitioner \quad-0.00
permittivity \quad-0.00\vglue-5pt
 \qquad $\vdots$\vglue-2pt}\vbox{\obeylines
point \quad-1.04
equation \quad-1.05
boundary \quad-1.06
social \quad-1.06
n \quad-1.09
relaxation \quad-1.14
fluid \quad-1.15
indian \quad-1.15
spin \quad-1.17
spin\_glass \quad-1.17
traffic \quad-1.18
system \quad-1.30
polymer \quad-1.33
class \quad-1.35
emerge \quad-1.36
gradient \quad-1.39
quantum \quad-1.43
surface \quad-1.43
synchronization \quad-1.45
market \quad-1.47
particle \quad-1.52
polyelectrolyte \quad-1.53
world \quad-1.57}
\hfill}\raise1.53in\hbox{$\left.\vbox{\vskip1.6in}\right)$}
\caption{Shown above are the most positive, a few intermediate, and most negative components of the q-bio classifying weight vector.}
\label{fig:intw}
\end{center}
\end{figure}

With a training set fully representative of the categories in question, it was then possible to run the classifier on the entirety of the same subject area content received from
1992 through Apr 2003, a total of 28,830 documents. The results are shown in
\fig{fig:qbio}. A total of 1649 q-bio documents was identified, and the trend towards an increasing percentage of q-bio activity among these arXiv users is evident: individual authors can be tracked as they migrate into the domain.  Visibility of the new domain can be further enhanced by referring submitters in real time, via the automated classifier running behind the submission interface.

Some components of the weight vector generated by the SVM for this training set are shown in \fig{fig:intw}.
Since the distance of a document to the classifying hyperplane is determined by taking the dot product with the normal vector, its component values can be interpreted as the classifying weight for the associated words.  The approach here illustrates one of the major lessons of the past decade: the surprising power of simple algorithms operating on large datasets. 

While implemented as a passive dissemination system, the arXiv has also played a social engineering role, with active research users developing an affinity to the system and adjusting their behavior accordingly.  
They scan new submissions on a daily basis, assume others in their field do so and are consequently aware of anything relevant that has appeared there (while anything that doesn't may as well not exist), 
and use it to stake intellectual priority claims in advance of journal publication.
We see further that machine learning tools can characterize a subdomain and thereby help accelerate its growth, via the interaction of an information resource with the social system of its practitioners.\footnote{Some other implications of these methods, including potential modification of the peer review system, are considered elsewhere \cite{Ginsparg/03}.} 

\medskip\noindent
{\bf Postscript: } The q-bio extraction described in section 5 was not 
just a thought experiment, but a prelude to an engineering experiment.
The new category went on-line in mid-September 2003,\footnote{See
http://arXiv.org/new/q-bio.html} at which time past submitters
flagged by the system were asked to confirm
the q-bio content of their submissions, and to send future submissions to the new category.
The activity levels at the outset corresponded precisely
to the predictions of the SVM text classifier, and later began to show indications
of growth catalyzed by the public existence of the new category.

\bibliographystyle{apalike}
\bibliography{sack03}

\end{document}